\documentstyle[12pt,epsf]{article}
\makeatletter
\def\submit#1{\gdef\@submit{\normalsize #1}}
\let\@null=\null
\gdef\null{%
    \@null\vskip -80pt\hbox{}\hfill
    \begin{tabular}[t]{r}
       \@date
    \end{tabular}
    \let\null=\@null
    \let\@date=\@submit}
\gdef\fnum@figure{\footnotesize Fig.~\thefigure}
\newif\iffig\figtrue
\long\def\testfig#1#2{\openin0 #1
    \ifeof0 \global\figfalse \message{(figure #1 not found)}
    \else\closein0 #2 \message{(including figure #1)} \fi}
\makeatother
\title{Theoretical
Expectations For High Mass Photon Pairs in $l^+ l^- \gamma \gamma$
Events at LEP/SLC\thanks{ Work supported in part by the U.S.  DOE
grant DE-FG05-91ER40627 and by the Polish Government grants KBN
203809101 and KBN 223729102.}}
\author{ \large S.  Jadach\thanks{
Permanent address:  Institute of Nuclear Physics, ul.\ Kawiory 26a,
Cracow, Poland}, B.F.L.  Ward and S.A.  Yost\\ \em Department of
Physics and Astronomy\\ \em University of Tennessee\\ \em Knoxville,
TN 37996-1200, USA}
\date{UTHEP-93-1002\\
      hep-ph/9402350\\
      October, 1993}
\submit{\hbox{}\hfill\\[1in] Submitted to {\em Physical Review} {\bf D}}
\def\figone%
{The process $e^+ e^- \rightarrow f\overline {f} + n(\gamma )$, with
$n(\gamma)$ radiation from both the initial and the final state. Here,
$f=e,\mu ,\tau ,u,d,s,c,b$ and $p_A$ is the four-momentum of fermion $A$.}
\def\figtwo%
{The ratio of the YFS3 di-photon mass distribution to the exact
$\mathop {\cal O}(\alpha ^2)$ mass distribution for both $e^+e^-\gamma
\gamma $ and $\mu ^+\mu ^-\gamma \gamma $ final states.}
\def\figthree%
{Results for $e^+e^-\gamma \gamma $ final states. Graph (a) shows the photon
mass distribution obtained using YFS3 and the exact $\mathop {\cal O}
(\alpha ^2)$ matrix elements. Graph (b) shows the same results normalized
to give the number of events expected for a total integrated luminosity
of 27 pb${}^{-1}$, together with a dotted line showing the YFS Monte Carlo
results used by the L3 Collaboration \protect\cite{1}.}
\def\figfour%
{Results for $\mu ^+\mu ^-\gamma \gamma $ final states. Graph (a) shows the
photon mass distribution obtained using YFS3 and the exact $\mathop {\cal O}
(\alpha ^2)$ matrix elements. Graph (b) shows the same results normalized
to give the number of events expected for a total integrated luminosity of
27 pb${}^{-1}$, together with a dotted line showing the YFS Monte Carlo results
used by the L3 Collaboration \protect\cite{1}.}
\begin{document}

\maketitle

\begin{abstract} Recently, the L3 collaboration has reported the
         observation of four events in the reactions $e^+ e^-
         \rightarrow l \overline{l} + \gamma\gamma$, $l = e,\mu,\tau$,
         with the invariant photon pair mass near 60 GeV in a data
         sample collected in the L3 detector corresponding to 950,000
         produced $Z^0$'s.  More recently, more data from the other
         LEP collaborations have become available.  In this paper, we
         use the Monte Carlo genrator YFS3 and our recent exact
         results on $e^+ e^- \rightarrow l \overline{l} +
         \gamma\gamma$ to assess the QED expectations for such L3-type
         high mass photon pair events in $e^+ e^- \rightarrow l
         \overline{l} + \gamma\gamma + n(\gamma)$ near the $Z^0$
         resonance.  \end{abstract} \newpage

\section{Introduction} Recently, the L3 collaboration reported
\cite{1} the observation of high-mass $\gamma\gamma$ pairs in the
reactions $e^+ e^- \rightarrow l \overline{l} + \gamma\gamma$ in the
$Z^0$ resonance region; in particular, four such events were reported
with the invariant $\gamma\gamma$ mass near 60 GeV from a data sample
corresponding to 950,000 produced $Z^0$'s in the LEP $e^+e^-$
colliding beam device.  More recently, related data from the ALEPH,
DELPHI, and OPAL Collaborations \cite{2} have become available, and
hence, an immediate issue which needs to be addressed is that of the
theoretical expectations from the basic QED processes themselves for
such high-mass photon pairs.  It is this issue which we shall address
in what follows.

More specifically, we want to use the YFS Monte Carlo approach
\cite{3} to higher-order $SU_{2L}\times U_1$ processes introduced by
two of us (S.J.and B.F.L.W.)  and the recent exact results \cite{4} by
the three of us on the processes $e^+ e^- \rightarrow l \overline{l} +
\gamma\gamma$,\ $l=e,\mu,\tau$, in the $Z^0$ resonance region to
assess the probability that the observations in Refs.\ \cite{1,2} are
consistent with higher-order QED processes.  This means that the
over-all normalization of our calculations in the L3 acceptance must
be known even in the presence of the strong initial state radiative
effects associated with the $Z^0$ resonance line shape.  Accordingly,
we will employ our YFS Monte Carlo event generator \cite{3}, which
treats $e^+ e^- \rightarrow f \overline{f} + n(\gamma)$, in the $Z^0$
resonance region with the $n(\gamma)$ multiple photon radiation for
both the initial and final fermion.  Here, $f$ is a fundamental
$SU_{2L}\times U_1$ fermion.  We should stress that, strictly
speaking, $f \ne e$ is implicit in YFS3.  However, in the L3
acceptance for the high mass $\gamma\gamma$ pairs, kinematical cuts
eliminate any large effect from the exchanges in $e^+ e^- \rightarrow
e^+ e^- + n(\gamma)$ which are not the $s$-channel $Z^0$ exchange, so
that we can use YFS3 for our analysis at the currently required level
of precision.  Our recent results in Ref.\ \cite{4} of course do not
have any such qualification in their applicability to the L3-type
events for $e^+ e^- \gamma\gamma$ final states:  the full matrix
element is available from Ref.\ \cite{4}, and indeed, it will be used
to check the validity of our YFS3 $s$-channel exchange approach to the
high $\gamma\gamma$ mass L3-type $e^+ e^- \gamma\gamma$ final states,
for example.

What we will do in this paper then is to set the QED higher order
expectations for the L3-type high $\gamma\gamma$ invariant mass
events.  We hope that sufficient data will be taken so that the
statistical errors on the experimental results analyzed in this paper
will cease to be the over-riding dominant error in the comparison
between theory and experiment.  We encourage the LEP/SLC
experimentalists to strive to accumulate the attendant factor $\sim
10$ in statistics required to reach this latter goal.

Our work is organized as follows.  In the next section, we present
some relevant theoretical and experimental background information.  In
Section 3, we compare our theoretical predictions with the LEP data.
In Section 4, we present some summarizing discussions.

\section{Preliminaries} The basic framework in which we shall work
will be that of the renormalization group improved YFS theory that is
realized via Monte Carlo methods via the event generators YFS2,
BHLUMI, and YFS3 in Refs.\ \cite{3}.  Since we shall focus on the YFS3
predictions for L3-type events, we begin by describing the relevant
aspects of the Monte Carlo realization of our YFS methods as they
relate to YFS3.

Specifically, for a process such as $e^+ e^- \rightarrow f
\overline{f} + n(\gamma)$, we have from Refs.\ \cite{5,6}, the
fundamental differential cross section \begin{eqnarray}
d\sigma_{{}_{\hbox{\tiny YFS}}} &=&
\exp\left\{2\alpha\mathop{\hbox{Re}} B + 2\alpha \widetilde B\right\}
\sum_{n=0}^\infty \int\prod_{j=1}^n {d^3 k_j\over k^0_j} \int {d^4
y\over (2\pi)^4}\ \nonumber\\* &\times& e^{iy(p_e + p_{\bar e} - p_f -
p_{\bar f} - \sum_j k_j) + D(y)} \;\overline\beta_n(k_1,\ldots,k_n)
{d^3 p_f d^3 p_{\bar f}\over p^0_f p^0_{\bar f}} \end{eqnarray} where
\begin{eqnarray} D(y) &=& \int{d^3 k\over k^0} \widetilde S(k)\;
\left(e^{-iyk} - \theta(k_{\hbox{\tiny max}} - k^0)\right), \\
2\alpha\widetilde B &=& \int_{k^0\le k_{{}_{\hbox{\tiny max}}}} {d^3
k\over k^0} \widetilde S(k), \\* B &=& {-i\over 8\pi^3} \int{d^4
k\over k^2 - m_\gamma^2 + i\epsilon} \nonumber\\*
&\times&\left[-\left({-2p_e - k \over k^2 + 2k\cdot p_e + i\epsilon} +
{-2p_{\bar e} + k \over k^2 - 2k\cdot p_{\bar e} + i\epsilon}
\right)^2 + \cdots\right], \end{eqnarray} with \begin{equation}
\widetilde S(k) = {\alpha\over 4\pi}\left[ -\left({p_{\bar e}\over k
\cdot p_{\bar e}} - {p_e \over k \cdot p_e}\right)^2 + \cdots\right].
\end{equation} Here, the kinematics is that illustrated in Fig.\ 1,
$m_\gamma$ is the photon infrared regulator mass, and $\bar\beta_n$
are the YFS hard photon residuals defined in Refs.\ \cite{5,6}, for
example.

\testfig{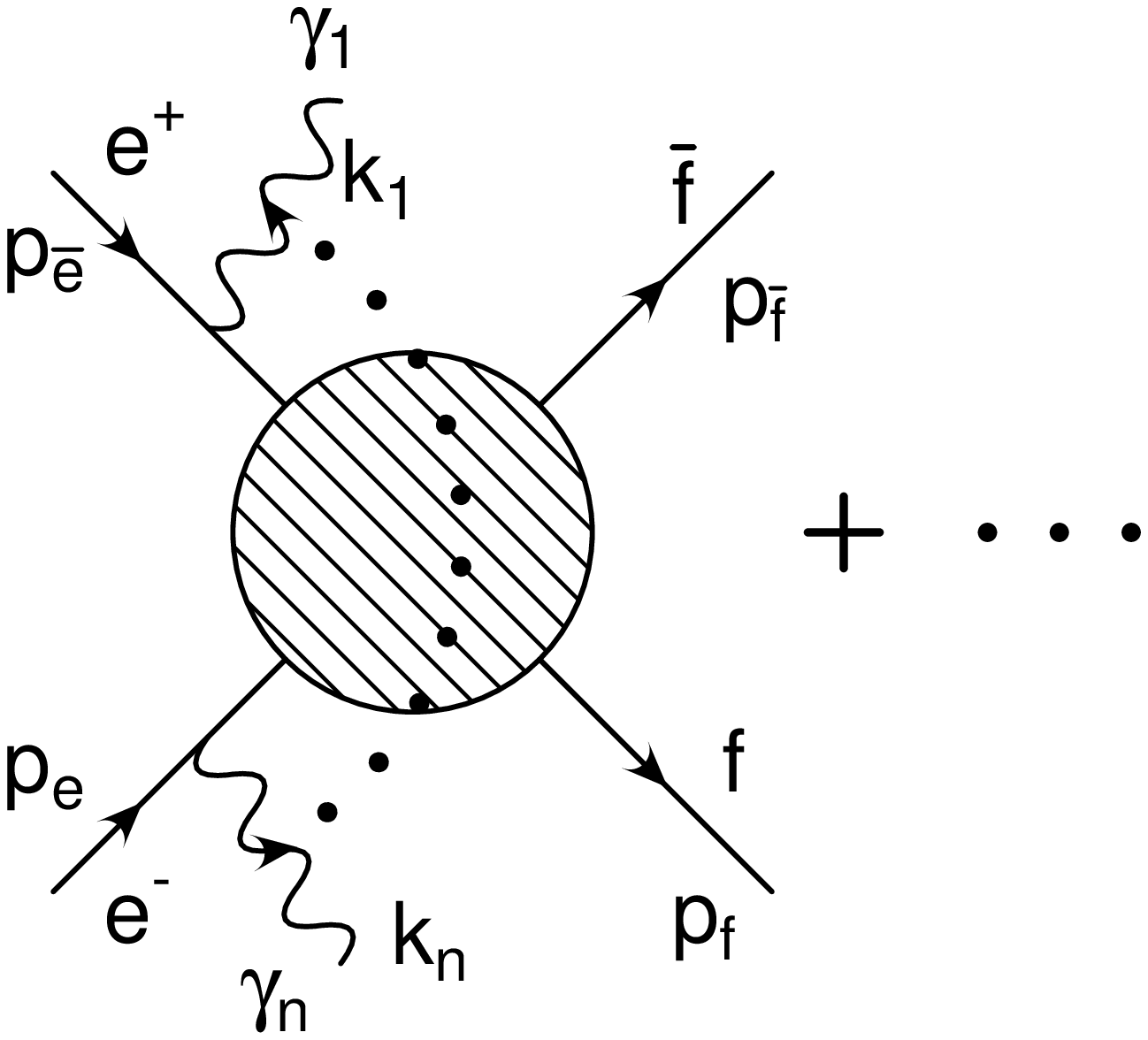}%
{
    \begin{figure}
        \epsfysize=2in
        \center
        \leavevmode
        \epsffile{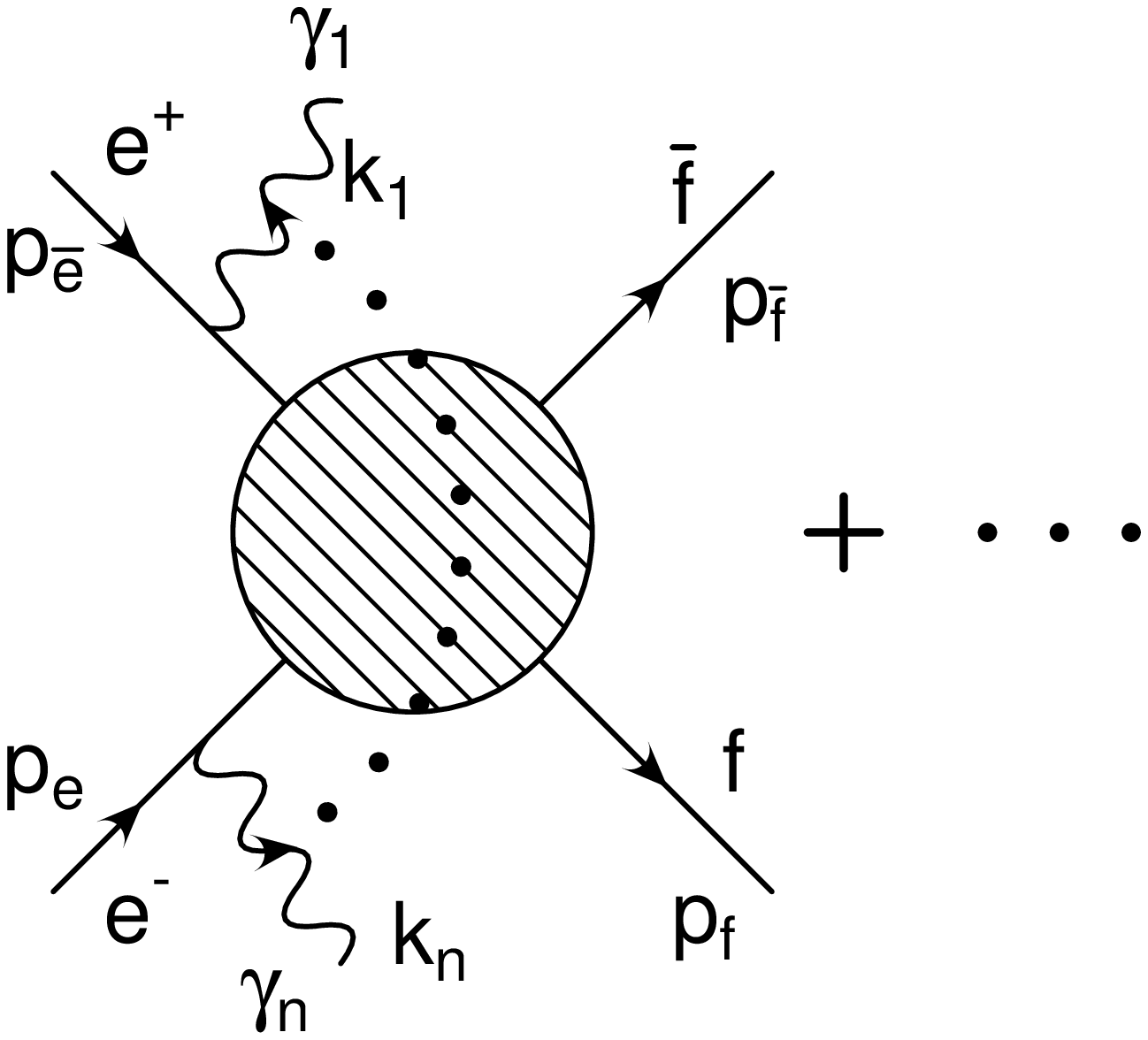}
        \caption{\figone}
    \end{figure}
}

In YFS3 \cite{3}, two of us (S.J.  and B.F.L.W.)  have realized (1)
via Monte Carlo methods for the case $f\ne e$ for both $n(\gamma)$
radiation from the initial state and $n(\gamma)$ radiation from the
final state, with the hard photon residuals $\overline\beta_{0,1,2}$
implemented in the respective MC to $\mathop{\cal O}(\alpha^2)$ at the
leading log level.  Thus, the $\mathop{\cal O}(\alpha)$ contributions
to $\overline\beta_{0,1,2}$ are exact and the $\mathop{\cal
O}(\alpha^2)$ contributions are correct to the leading log level.  It
follows that, in a special region of the event phase space for events
of the L3-type, it is necessary to check that the leading-log
$\mathop{\cal O}(\alpha^2)$ approximation for the respective
hard-photon effects is indeed accurate to the desired level of
accuracy.  It is this issue that we discuss in our following analysis.

\section{Comparison of YFS3 and Exact $\mathop{\cal O}(\alpha^2)$
Results for L3-Type Events}

In this section, we compare the exact $\mathop{\cal O}(\alpha^2)$
results in Ref.\ \cite{4} for the process $e^+ e^- \rightarrow l
\overline l + 2\gamma$, restricted to the L3-type phase space cuts as
they are given in Ref.\ \cite{1}, with those predicted by YFS3.  Here,
we emphasize immediately that, due to the wide angles of the photons
with respect to the charged particles, the effect of radiative
corrections on the $Z^0$ line shape has been taken into account
properly in YFS3, and this amounts to an over-all normalization
correction to the cross section for L3-type events.  Upon removing
this $Z^0$ line shape effect, we are left with a comparison of the
YFS3 two hard photon distributions in the L3-type events phase space
with the analogous distributions as given by the exact $\mathop{\cal
O}(\alpha^2)$ result.  It is this comparison which we now present.

\testfig{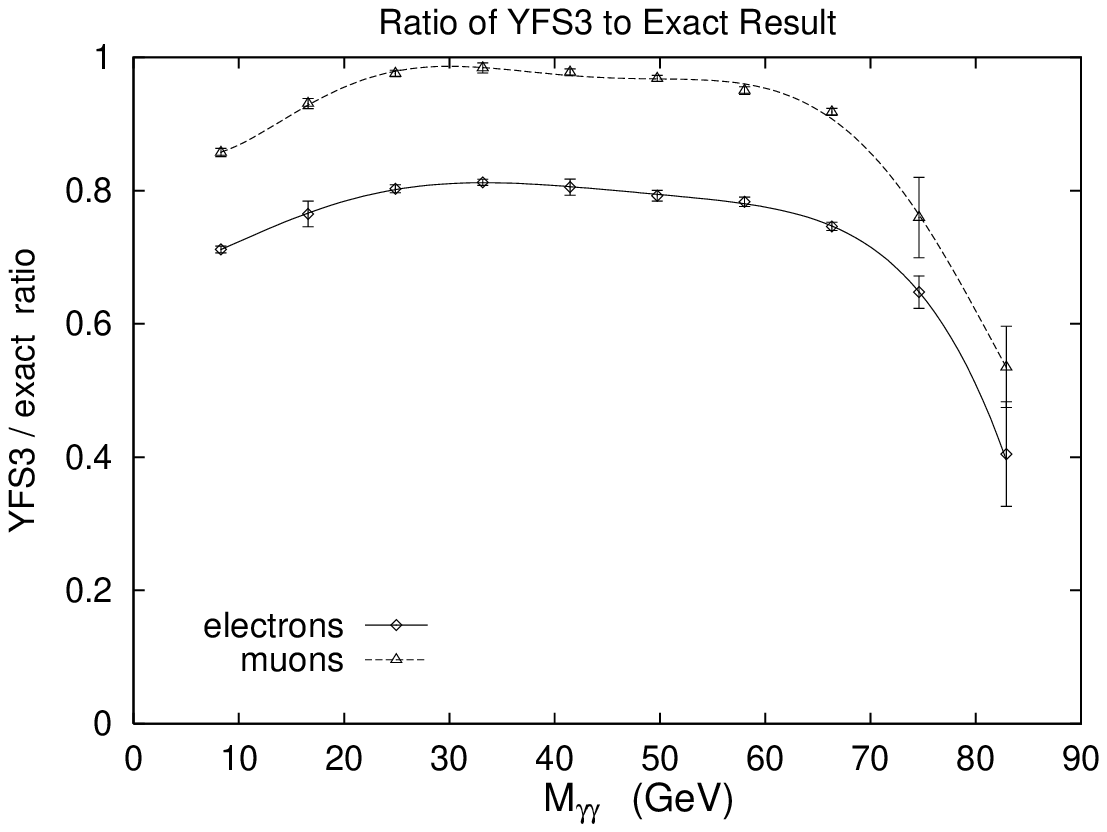}%
{
    \begin{figure}
        \center
        \leavevmode
        \epsfysize=3in
        \epsffile{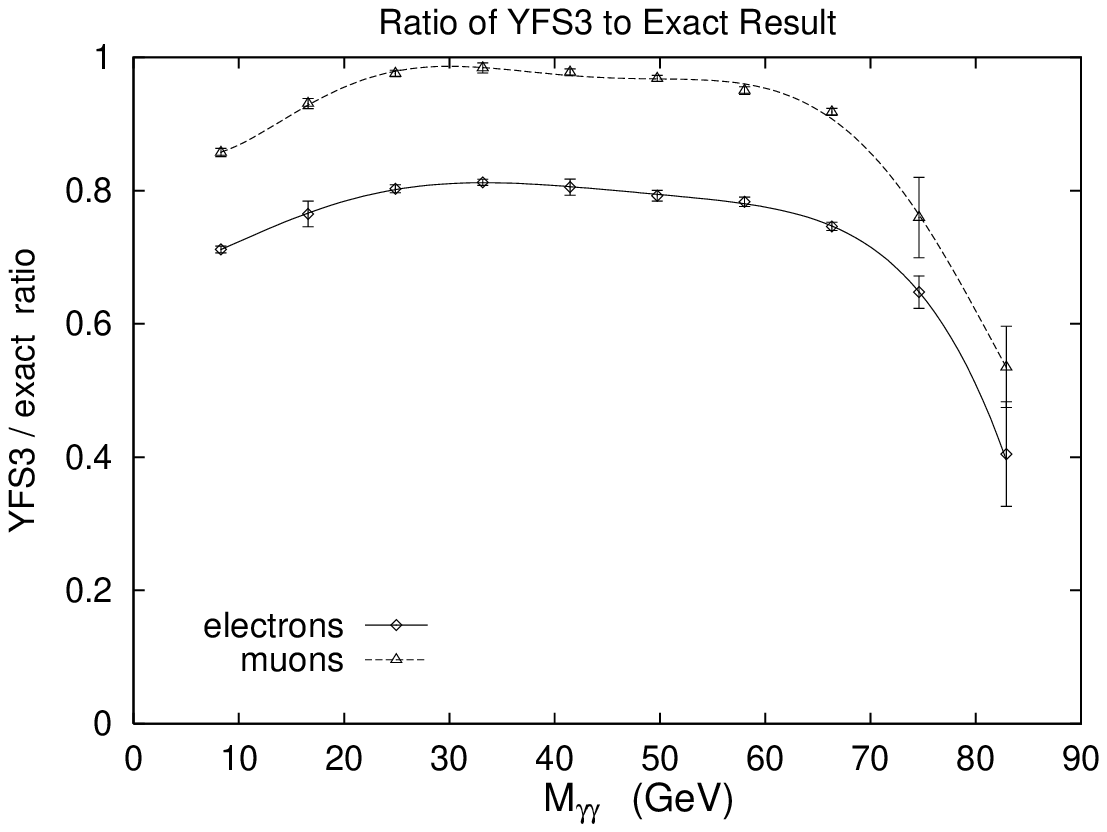}
        \caption{\figtwo}
    \end{figure}
}

Specifically, we focus on the ratio of the YFS3 two hard photon
leading log matrix element squared and the exact $\mathop{\cal
O}(\alpha^2)$ two hard photon matrix element squared in the L3-type
events phase space by plotting this ratio as a function of
$M_{\gamma\gamma}$, the respective di-photon invariant mass, in such
L3-type events.  This we do in Fig.\ 2 for the cases $e^+ e^-
\rightarrow e^+ e^- + 2\gamma$ and $e^+ e^- \rightarrow \mu^+\mu^- +
2\gamma$.  We see that, indeed, our YFS3 matrix element squared is
within $78\%$ of the exact $\mathop{\cal O}(\alpha^2)$ result for the
$e^+ e^- \rightarrow e^+ e^- + 2\gamma$ case in the 60 GeV regime, and
is within $94\%$ of the exact $\mathop{\cal O}(\alpha^2)$ result in
the $e^+ e^- \rightarrow \mu^+ \mu^- + 2\gamma$ case.  Alternatively,
we plot in Figs.  3 and 4 (for the $e^+e^-\gamma\gamma$ and
$\mu^+\mu^-\gamma\gamma$ cases respectively) the di-photon mass
distribution in the L3-event phase space for the exact $\mathop{\cal
O}(\alpha^2)$ and YFS3 two hard photon matrix elements squared, both
(a) as $d\sigma/dM_{\gamma\gamma}$, and (b) as a histogram of the
number of expected events versus $M_{\gamma\gamma}$ for 27 pb${}^{-1}$
of integrated luminosity.  Also shown in Figs.\ 3(b) and 4(b) are the
MC results from the L3 paper, Ref.\ \cite{1}, which are just the YFS3
results, of course.  We conclude that, in all cases in Figs.\ 3 and 4,
there is good agreement between our exact $\mathop{\cal O}(\alpha^2)$
expectations and those results generated by YFS3.

\testfig{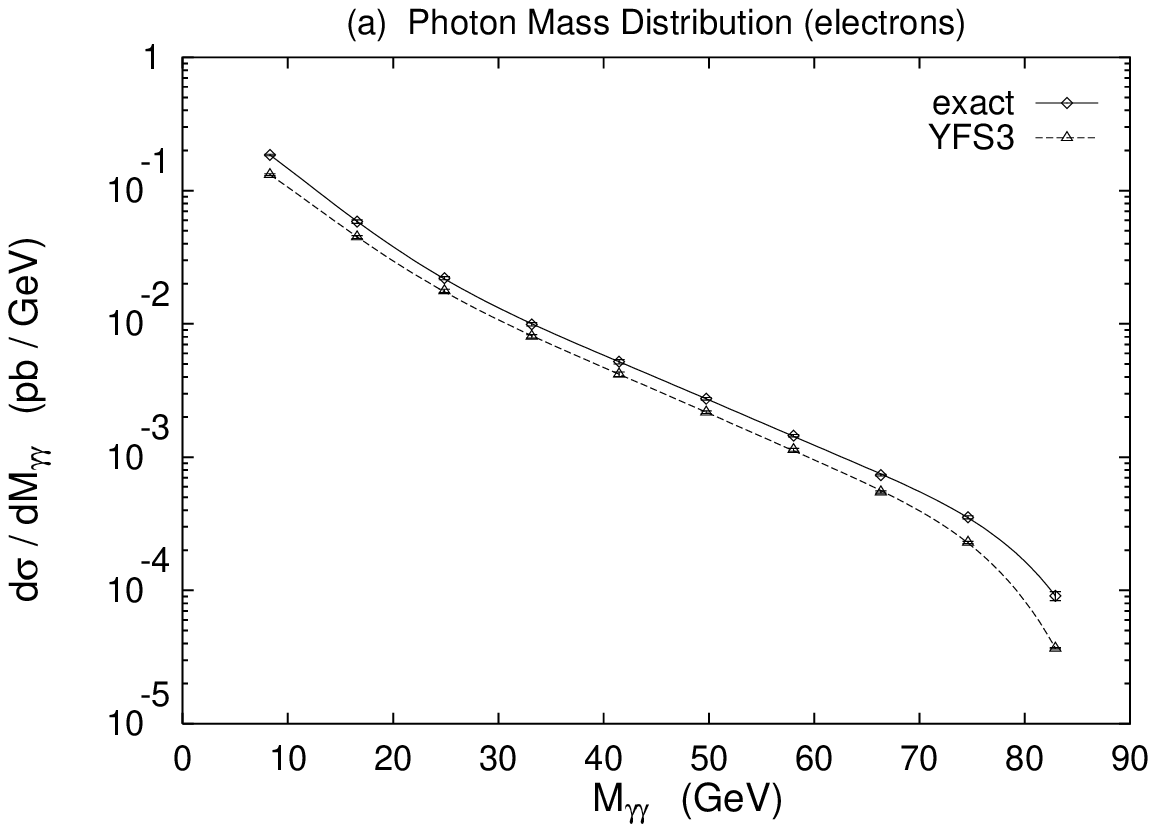}%
{
    \testfig{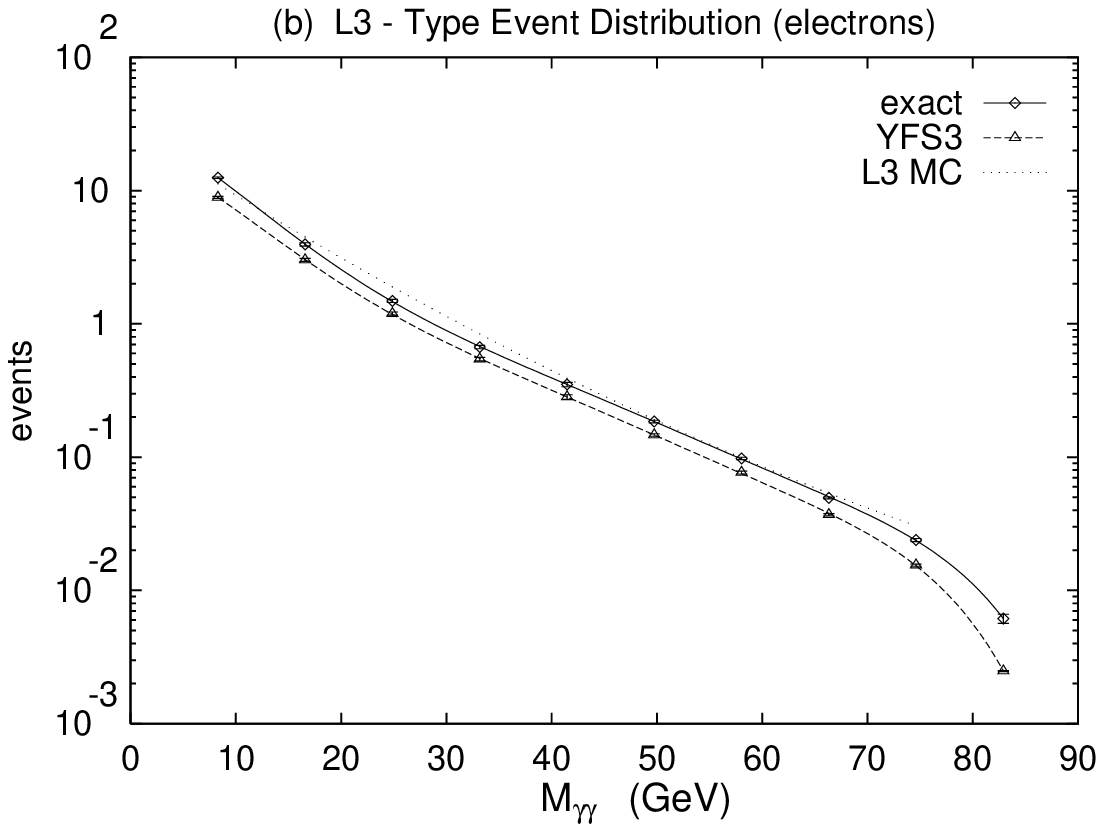}%
    {
        \begin{figure}
            \center
            \leavevmode
            \epsfysize=3in
            \epsffile{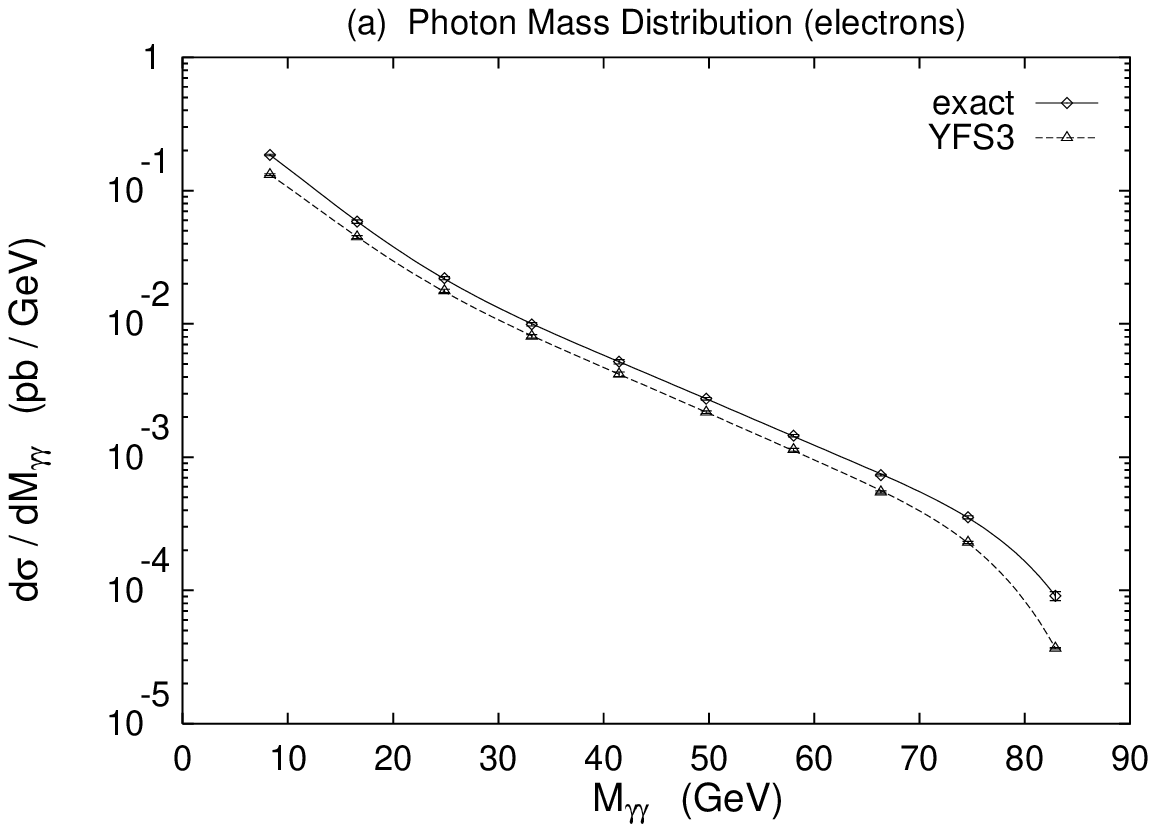}
            \par
            \center
            \leavevmode
            \epsfysize=3in
            \epsffile{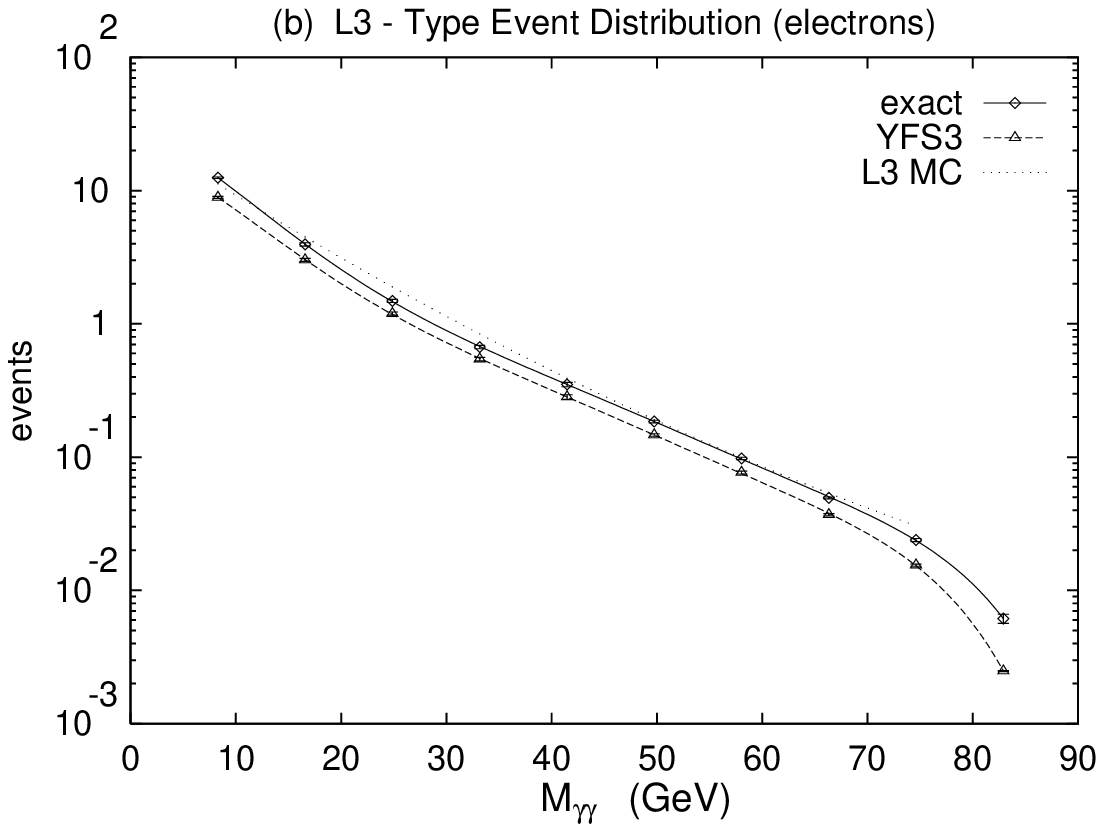}
            \caption{\figthree}
        \end{figure}
    }
}

\testfig{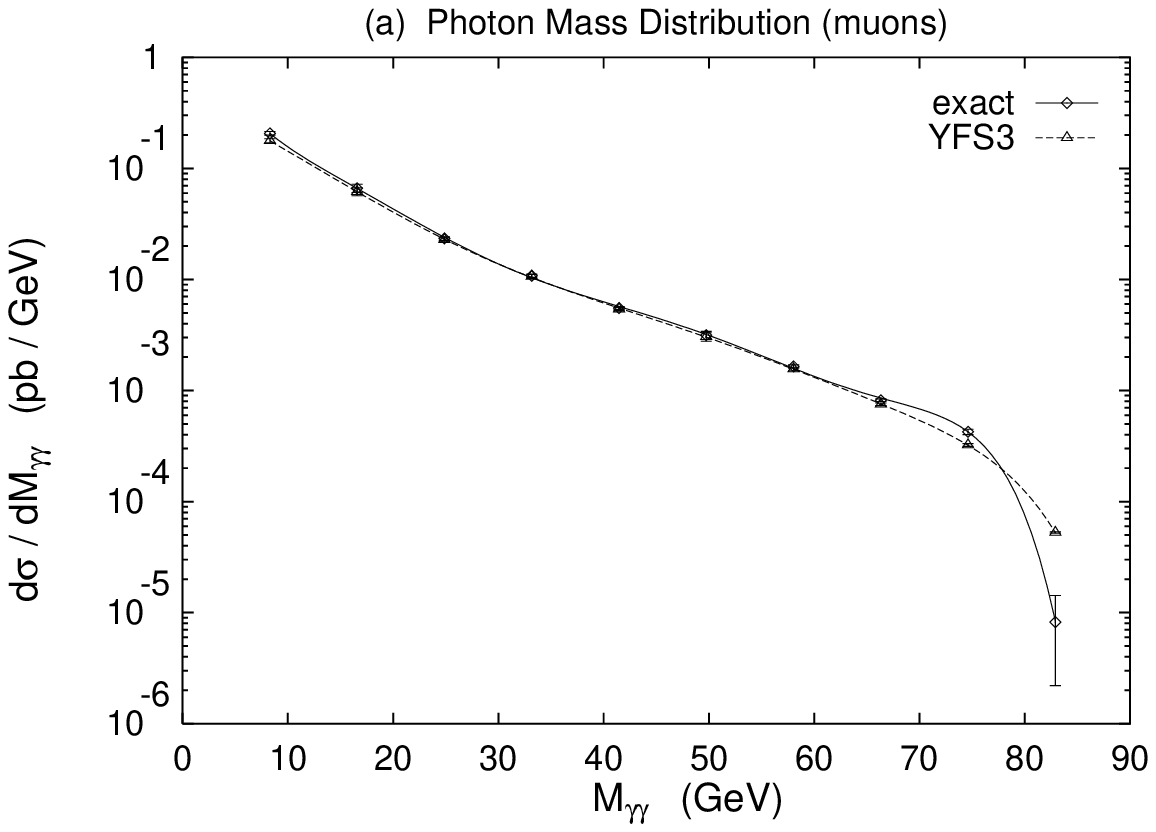}%
{
    \testfig{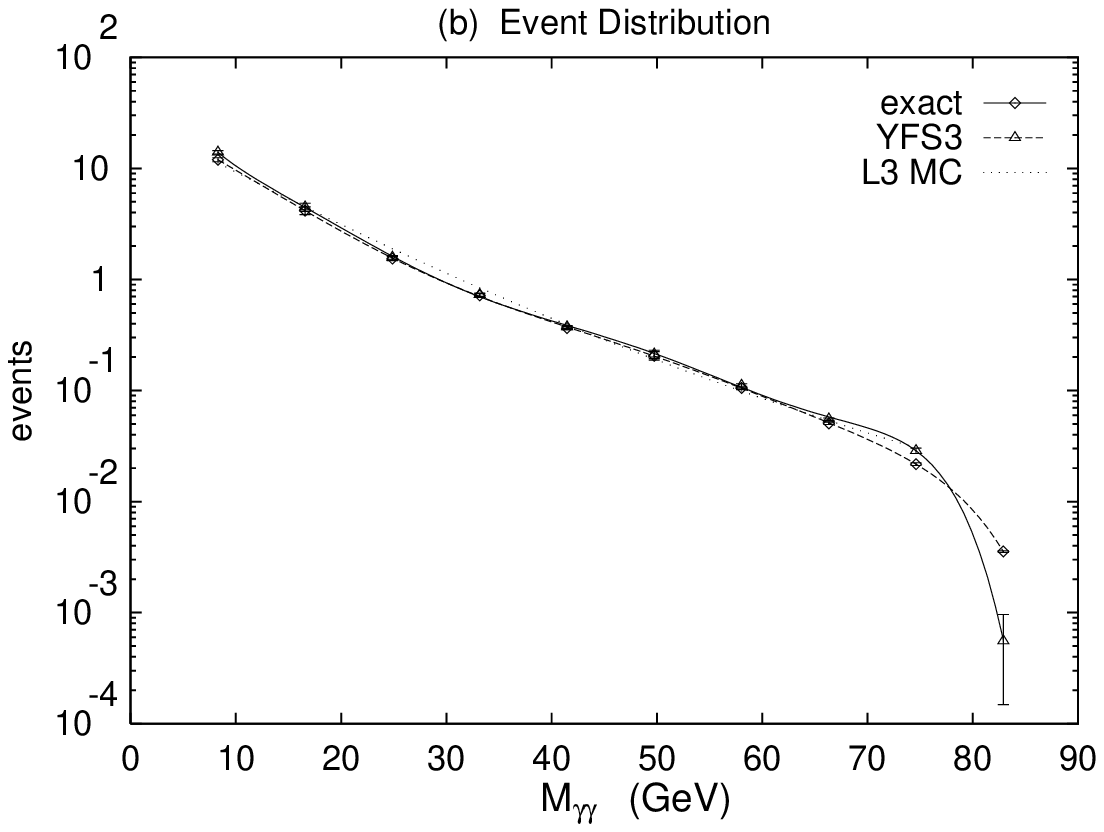}%
    {
        \begin{figure}
            \center
            \leavevmode
            \epsfysize=3in
            \epsffile{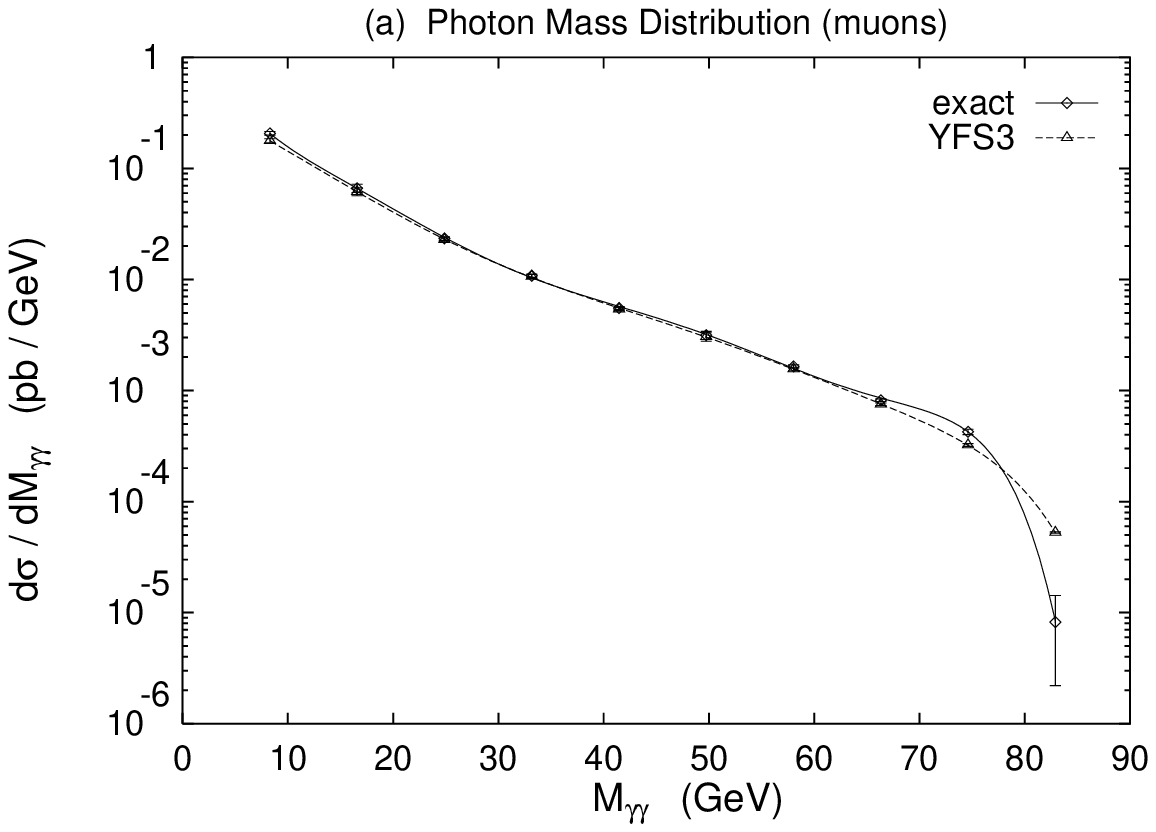}
            \par
            \center
            \leavevmode
            \epsfysize=3in
            \epsffile{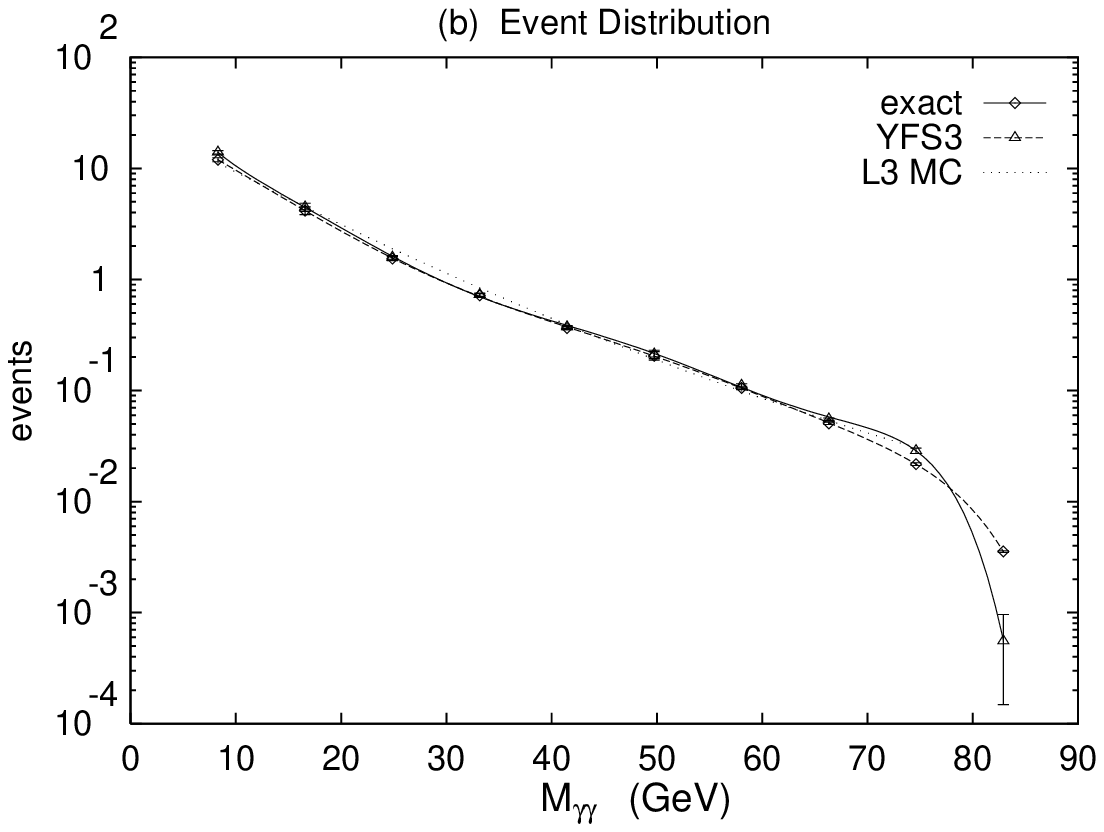}
            \caption{\figfour}
            \end{figure}
    }
}

Recently, all LEP collaborations have searched for L3-type events.
The results of their search are discussed in Ref.\ \cite{2}.  Here, we
note that, in the regime of $M_{\gamma\gamma}$ between 50 GeV and 80
GeV, the LEP collaborations find 15 events in the L3-type $2\gamma$
phase space, and YFS3 predicts 9.  The probability that this is a
statistical fluctuation is $1.9\%$.  What we can say here is that the
statistical effects in the YFS3 comparison with data is indeed the
dominant source of uncertainty in that comparison.  Thus, we urge the
experimentalists to strive for more data so that the nature of these
observations can be clarified.

\section{Conclusions} We have analyzed the L3-type
$l\overline{l}\gamma\gamma$ event high di-photon mass spectrum in
YFS3, the second order leading log YFS-exponentiated final + initial
state $n(\gamma)$ radiation event generator, in comparison to the
exact $\mathop{\cal O}(\alpha^2)$ prediction as determined by the
results in Ref.\ \cite{4}.  We find good agreement between these two
independent calculations of the respective spectra.  This agreement
means that the use of YFS3 to estimate the probability that the
observed L3-type events at LEP are a QED fluctuation does not suffer
from an unknown physical precision error associated with its use of
$\mathop{\cal O}(\alpha^2)$ leading-log matrix elements for hard
2-photon emission into the respective L3-type
$l\overline{l}\gamma\gamma$ phase space.

We want to note that a comparison of YFS3 with the $\mathop{\cal O}
(\alpha^2)$ exact results in Ref.\ \cite{7} has also been carried out
\cite{1,8} and it agrees with our findings.  The way is open to
incorporate the exact $\mathop{\cal O}(\alpha^2)$ matrix element for
the L3-type $l\overline{l}\gamma\gamma$ event phase space into YFS3 if
the statistics on these events would be increased to require such
accuracy in the YFS3 predictions.  We encourage the LEP
experimentalists to strive for such an increase in L3-type
$l\overline{l}\gamma\gamma$ event statistics.

\section*{Acknowledgments} Two of the authors (S.J.  and B.F.L.W.)
thank Prof.\ J.  Ellis for the kind hospitality and support of the
CERN TH Division, where a part of this work was completed.  The
authors have also benefitted from discussions with J.  Qian, K.
Riles, E.  R.-W\accent'30 as, Z.  W\accent'30 as and B.  Wyslouch.

\iffig\relax\else
\newpage
\section*{List of Figures}
\begin{trivlist}
    \item[Fig.\ 1: ] \figone
    \item[Fig.\ 2: ] \figtwo
    \item[Fig.\ 3: ] \figthree
    \item[Fig.\ 4: ] \figfour
\end{trivlist}
\fi

\end{document}